\documentclass[prd,aps,superscriptaddress,twocolumn,floatfix,nofootinbib]{revtex4}
\pdfoutput=1

\usepackage{amsfonts}
\usepackage{amsmath}
\usepackage{amssymb}
\usepackage{bm}
\usepackage{dcolumn}
\usepackage{graphicx}   
\usepackage[latin1]{inputenc}
\usepackage{latexsym}
\usepackage{rotating}
\usepackage{hyperref}
\usepackage{graphicx}
\usepackage{color}
\usepackage{subfigure}

\usepackage{soul}
\newcommand\be{\begin{equation}}
\newcommand\ba{\begin{eqnarray}}
\newcommand\ee{\end{equation}}
\newcommand\ea{\end{eqnarray}}

\begin{document}

\title{Graviton to Photon Conversion via Parametric Resonance}
  
\author{Robert Brandenberger}
\email{rhb@physics.mcgill.ca}
\affiliation{Department of Physics, McGill University, Montr\'{e}al,
  QC, H3A 2T8, Canada}

\author{Paola C. M. Delgado}
\email{paola.moreira.delgado@doctoral.uj.edu.pl}
\affiliation{Faculty of Physics, Astronomy and Applied Computer Science, Jagiellonian University, 30-348 Krakow, Poland}

\author{Alexander Ganz}
\email{alexander.ganz@uj.edu.pl}
\affiliation{Faculty of Physics, Astronomy and Applied Computer Science, Jagiellonian University, 30-348 Krakow, Poland}
 
\author{Chunshan Lin}
\email{chunshan.lin@uj.edu.pl}
\affiliation{Faculty of Physics, Astronomy and Applied Computer Science, Jagiellonian University, 30-348 Krakow, Poland}


\begin{abstract}

We study the parametric resonance excitation of the electromagnetic field by a gravitational wave. We show that there is narrow band resonance. For an electromagnetic field in the vacuum the resonance occurs only in the second band, and its strength is thus suppressed by two powers of amplitude of the gravitational wave. On the other hand, in the case of an electromagnetic field in a medium with the speed of light smaller than 1 (in natural units), there is a band of Fourier modes which undergo resonance in the first band. 
 
\end{abstract}

\maketitle

\section{ Introduction~~~}
Parametric resonance is a well-known effect in classical mechanics: an oscillator with a periodically varying contribution to the mass will be exponentially excited if the frequency of the oscillator lies in certain resonance bands determined by the frequency of the variation of the mass (see e.g. \cite{Landau, Arnold} for textbook treatments). If the amplitude of the varying part of the mass is small compared to the magnitude of the time-independent part, we speak of ``narrow band resonance'', if it is large then we are in the realm of ``broad resonance''. Parametric resonance is a special case of the Floquet theory of instability of a dynamical system in the presence of a periodic time-dependence of one of the coefficients \cite{Floquet}. The equation of motion for the special case is called the ``Mathieu equation''.

In early univere cosmology, parametric resonance plays a crucial role in the transfer of energy to regular matter at the end of a hypothetical period of inflation \cite{TB, DK}. At the end of inflation, the scalar field $\phi$ which drives inflation will be oscillating about the minimum of its potential. This oscillation can induce a parametric resonance instability for any field $\chi$ which couples in an appropriate way to $\phi$, e.g. via a $\phi^2 \chi^2$ coupling in the case of a matter scalar field $\chi$. In the case of a self-interacting scalar field $\phi$, excitation of fluctuations of $\phi$ will also occur (see \cite{RHrev, Karouby} for reviews). This instability is known as ``preheating'' \cite{KLS1, STB, KLS2}.
Note that the preheating instability can occur for both bosons and fermions, although because of Pauli blocking the resonance for fermions is less efficient \cite{Felder}.

In an expanding universe the equation of motion for a matter field $\chi$ contains a Hubble damping term, and hence the parametric resonance analysis does not directly apply. However, if we rescale the matter field by a power of the cosmological scale factor and also work in conformal time $\tau$ instead of physical time $t$, we obtain an equation of motion without damping term. However, the bare mass term of the equation in terms of the original field now acquires a scale factor dependence which greatly reduces the efficiency of the Floquet resonance. On the other hand, for massless fields we obtain a standard Mathieu equation for the rescaled field.

Gravitational waves induce oscillating terms in the equations of motion for all matter fields. In the case of massless matter fields such as the photon, it is hence expected that these gravitational wave can induce instabilities. These instabilities, in turn, will drain energy from the gravitational waves. In the Standard Model, the only massless field is the photon \footnote{If there is a massless neutrino there can also be an instability to neutrino production, but because of Pauli blocking it will be less efficient than for photons.}. Here, we will study the parametric resonance instability of the photon field in the presence of a gravitational wave. We find that there is indeed a resonance effect. In vacuum, the resonance occurs only in the second resonance band and is hence highly inefficient. On the other hand, in a medium in which the speed of fluctuations of the electromagnetic field is smaller than unity \footnote{We use units in which the speed of light, Planck's constant and Boltzmann's constant are $1$.}, the instability occurs in the first resonance band and is hence much more efficient. In the current work, we estimate the decay rate of a packet of gravitational waves passing through a medium. More details will be given in a follow-up paper.
 
In the following we will be studying the effects of gravitational waves on matter fields in a Minkowski space-time backgound. Provided that the
time scale of the instability is shorter than the duration which a
mode spends in the instability band, the effects of the expansion of space are small, the main effect being that modes slowly enter and exit the resonance bands, as argued already in the original article \cite{TB}. Note that our analysis does not make use of any physics beyond Standard Model particle physics and Einstein gravity.\\

\section{ Massless Scalar Field Resonance~~~} 
Here we consider a gravitational wave of frequency $\omega$ exciting a scalar field $\phi$ with mass $m_{\phi}$. We consider a gravitational wave with metric tensor $h_{\mu\nu}$ travelling in Minkowski space-time. The full metric is
\be
g_{\mu\nu}=\eta_{\mu\nu}+h_{\mu\nu}.
\ee
Specifically, we consider a standing gravitational wave with frequency $\omega$:
\be \label{standingGW}
h_{ij} = h_0\cos\omega t\cos\omega z\cdot\epsilon_{ij},\qquad h_{0\mu}=0,
\ee
where $h_0$ is the amplitude, and $\epsilon_{ij}$ is the polarisation tensor
\be
\epsilon_{ij} =\left(\begin{array}{ccc}
1 & 1 & 0\\
1 & -1 & 0\\
0 & 0 & 0
\end{array}\right).
\ee

The equation of motion of a scalar field of mass $m$ in this gravitational wave background is
\be
\ddot{\phi}-\left(\delta_{ij}-h_{ij}\right)\partial_i\partial_j\phi+m_{\phi}^2\phi=0.
\ee
This equation is reminiscent of the sound speed resonance mechanism \cite{Cai:2018tuh}\cite{Cai:2020ovp}, where the sound speed of scalar modes or tensor modes receives an oscillatory correction which eventually triggers the resonance instability.  
In Fourier space, the equation becomes
\be
\ddot{\phi}_{\textbf{k}}+\left(k^2+m_{\phi}^2\right)\phi_{\textbf{k}} -\frac{h_0}{2}k_\epsilon^2\cos\omega t\cdot\left[\phi_{\textbf{k}-\textbf{p}}+\phi_{\textbf{k}+\textbf{p}}\right]=0,
\ee
where $k_\epsilon^2\equiv \epsilon_{ij}k_ik_j$ and $\textbf{p}$ is a 3-dimensional vector defined by $\textbf{p}=(0,0,\omega)$.  Let us define the variable
\be
\Phi\left(t,k_x,k_y,k_z\right)\equiv\phi\left(t,k_x,k_y,k_z\right)+\phi\left(t,k_x,k_y,-k_z\right),
\ee
and let us choose $k_z = \omega/2$. Then, quite remarkably, its equation of motion is
\be \label{eomPhi}
\ddot{\Phi}_{k_z}+\left(k^2+m_{\phi}^2\right)\Phi_{k_z}-\frac{h_0}{2}k_\epsilon^2\cos\omega t\left(\Phi_{k_z}+\Phi_{3k_z}\right)=0,
\ee
which is a Mathieu equation with a source term proportional to $\Phi_{3k_z}$. To avoid notational clutter we have defined $\Phi_{k_z}\equiv\Phi\left(t,k_x,k_y,k_z\right)$. In a first approximation, the source term can be neglected as the mode  $\Phi\left(t,k_x,k_y,3k_z\right)$ does not receive a parametric resonance amplification and thus remains small \footnote{We can include the $\Phi\left(t,k_x,k_y,3k_z\right)$ term and add in the equation of motion for this mode, thus obtaining a set of coupled differential equations. In the context of a study of the effects of inhomogeneous noise on the strength of parametric resonance, it has been shown that considering the inhomogeneous system actually boosts the growth rate of the instability \cite{Craig2}. This is an consequence of Furstenberg's Theorem \cite{Furstenberg} (see \cite{Craig1}). As an application, this leads to a new proof of Anderson Localization in condensed matter systems \cite{Craig3}.}. 

Inserting the value of $k_z$, 
the equation of motion (\ref{eomPhi}) then becomes the standard Mathieu equation
\be \label{Mathieu}
{\Phi_{k_z}}^{\prime\prime} + \bigl[A_k - 2q {\rm{cos}}(2\tau) \bigr] {\Phi_{k_z}} \, = \, 0 \, ,
\ee
 with the rescaled time variable being $\tau \equiv \frac{\omega t}{2}$, the prime denotes the derivative with respect to $\tau$,  and
\begin{eqnarray}
A=1+\frac{4\left(k_x^2+k_y^2+m_\phi^2\right)}{\omega^2},\qquad
q=\frac{h_0k_\epsilon^2}{\omega^2}\,.
\end{eqnarray}
The Mathieu equation (\ref{Mathieu}) undergoes broad resonance for $q>1$, where the exponential instability occurs for all sufficiently long wavelength modes, and narrow resonance for $q<1$, where the exponential instability occurs only for narrow  bands of $k$ modes.
We are interested in the weak field limit where the amplitudes of both polarisations are small, i.e. $h_0\ll1$. Hence, $q \ll 1$ and we are dealing with narrow band resonance. It is clear that we are outside of the first resonance band where $A\subset\left(1-q,1+q\right)$. However,  parametric resonance may still occur at the second resonance band where $A\subset\left(4-q^2,4+q^2\right)$.  For resonance in the second band, the amplitude of $\Phi$ grows as $\exp[\mu_k^{(2)}\tau]$, where $\mu_k^{(2)}$ is the Floquet exponent of the second resonance band $\mu_k^{(2)}\simeq \frac{q^2}{4},$
which is parametrically suppressed compared to the Floquet exponent in the first resonance band which
 is $\mu_k^{(1)} \simeq \frac{q}{2}$.

There is, however, a way to obtain resonance in the first band: if we consider the propagation of the scalar field $\phi$ in a medium which leads to a reduced speed of propagation $c_s < 1$, then $A$ becomes
\be
A \, = c_s^2 + c_s^2\frac{4(k_x^2 + k_y^2)}{\omega^2} + \frac{4m_{\phi}^2}{\omega^2} \, .
\ee
In this case, for a massive scalar field it will remain impossible to obtain resonance in the first band, unless $m_{\phi}^2 < (1-c_s^2) \omega^2/4$, which in our case is not reasonable for masses of Standard Model particles, given that the wavelengths of gravitational waves emitted by the most of astrophysical events are of macroscopic scale. 
However, for photons (which are massless) there will be a band of $(k_x, k_y)$ values which lie in the first resonance band. Thus, in the following we will focus on gravitational waves exciting the electromagnetic field.

One may be confused at this point, as quantum field theory tells us a massless particle does not decay to a massive particle in the vacuum; The process is simply forbidden by energy momentum conservation. Nevertheless, two colliding massless particles do decay to massive particles, as now this process is allowed (for instance a pair of colliding high-energy photons can decay into an electron-positron pair). This is precisely the case in our analysis, where a standing gravitational wave, which can be understood as the collective behaviour of two groups of massless gravitons travelling in the opposite direction, decays into massive scalar particles (provided that mass is smaller than the frequency of gravitational wave) due to the collision of massless gravitons.

In passing, we shall mention that for a traveling gravitational wave in  vacuum, the parametric resonance does not occur, even if the scalar field is massless. This is because the lightcones of the gravitational wave and the scalar field overlap with each other. Sitting on the wavefront of the scalar wave, one does not ``feel" the oscillation induced by the gravitational wave. However, in a medium where the scalar wave is sub-luminal, the two lightcones do not overlap and that opens up the channel converting energy from the gravitational wave sector to the scalar field sector, even for a pure traveling wave. The similar effect has  been observed in the framework of the modified gravity too \cite{Creminelli:2019nok}. More details will be covered in our followup paper \cite{us}. \\

 
 \section{ Electromagnetic Resonance~~~} 
Here we consider the excitation of the electromagnetic field by a gravitational wave in a medium with speed of light $c_s < 1$. The metric which enters the kinetic part of gauge field equation of motion is
\be
g_{\mu\nu}=\tilde{\eta}_{\mu\nu}+h_{\mu\nu},
\ee
where $\tilde{\eta}_{\mu\nu}=(-1,1/c_s^2,1/c_s^2,1/c_s^2)$. Generally $c_s^2$ is dependent  on the frequency. In our idealised case where a mono-frequency  gravitational wave is considered, $c_s^2$ is just a constant. It can be a good approximation if the frequency spread in a wave packet is small. 

The Coulomb gauge is unavailable in the presence of the gravitational wave, and thus we adopt the Weyl gauge instead, where $A_0=0$. The $i-th$ component of the equation of motion reads
\ba \label{eomAi}
0&=&g_{i\alpha}\partial_\mu\left(F_{\rho\sigma}g^{\alpha\rho}g^{\mu\sigma}\right)
\nonumber \\
&=&\partial_t^2 A_i-c_s^2\partial_t h_{ij}\cdot\partial_tA_j+c_s^2\partial_jF_{ij} \nonumber \\
& & - c_s^4h_{jk}\partial_j F_{ik}-c_s^4F_{kj}\partial_{j}h_{ik} \, ,
\ea
and the $0-th$ component gives the modified  Gauss law,
\be
\partial_i E_i=c_s^4h_{ij}\partial_iE_j.
\ee

We consider an unpolarised standing gravitational wave
\be
h_{ij}=h_0\cos\omega t\cos\omega z\cdot\epsilon_{ij} \, .
\ee
The generalizations to other waves and other types of polarisation are straightforward, and will be covered in our followup paper \cite{us}. 

Translating Eq. (\ref{eomAi}) to momentum space and defining
\ba
\mathcal{A}_x\left(t,k_x,k_y,k_z\right) &\equiv& A_x\left(t,k_x,k_y,k_z\right)+A_x\left(t,k_x,k_y,-k_z\right),\nonumber\\
\mathcal{A}_y\left(t,k_x,k_y,k_z\right) &\equiv& A_y\left(t,k_x,k_y,k_z\right)+A_y\left(t,k_x,k_y,-k_z\right),\nonumber\\
\mathcal{A}_z\left(t,k_x,k_y,k_z\right) &\equiv& A_z\left(t,k_x,k_y,k_z\right)-A_z\left(t,k_x,k_y,-k_z\right),\nonumber\\
\ea
then for $k_z=\omega/2$ these equations (\ref{eomAi}) can be written in matrix form
\be\label{eomAi1}
\ddot{\mathcal{Y}}+c_s^2\mathcal{G}\mathcal{Y}+c_s^2\mathcal{F}\dot{\mathcal{Y}}+c_s^4\mathcal{M}\mathcal{Y}\simeq 0,
\ee
where 
\be
\mathcal{Y}=\left(\begin{array}{ccc}
\mathcal{A}_x\\
\mathcal{A}_y\\
\mathcal{A}_z\\
\end{array}\right)
\ee
$\mathcal{G}$ is the gradient matrix
\be
\mathcal{G}=\left(\begin{array}{ccc}
k_y^2+\frac{\omega^2}{4}~&-k_xk_y~&-k_xk_z\\
-k_xk_y~&k_x^2+\frac{\omega^2}{4}~&-k_yk_z\\
-k_xk_z~&-k_yk_z~&k_x^2+k_y^2\\
\end{array}\right)
\ee
$\mathcal{F}$ is the friction matrix
\be
\mathcal{F}=\frac{1}{2}h_0\omega\sin\omega t\left(\begin{array}{ccc}
1&1&0\\
1&-1&0\\
0&0&0\\
\end{array}\right),
\ee
and $\mathcal{M}$ is defined by
\begin{widetext}
\be
\mathcal{M}=\frac{1}{4}h_0\cos\omega t\left(\begin{array}{ccc}
-2k_x k_y+2k_y^2+\omega^2~&2k_x^2-2k_xk_y+\omega^2~&-2\omega\left(k_x+k_y\right)\\
2k_x k_y+2k_y^2+\omega^2~&-2k_x^2-2k_xk_y-\omega^2~&-2\omega\left(k_x-k_y\right)\\
-\left(k_x+k_y\right)\omega~&\left(k_y-k_x\right)\omega~&2k_\epsilon^2\\
\end{array}\right).
\ee
\end{widetext}
Note of that only two of the variables are independent since the photon has only two dynamical degrees of freedom. Thus we need to decouple one of variables from the other two. The gradient matrix $\mathcal{G}$ has only two non-vanishing eigenvalues, 
\be
\mathcal{S}\mathcal{G}\mathcal{S}^{-1}=\left(\begin{array}{ccc}
0&0~&0\\
0~&k_x^2+k_y^2+k_z^2~&0\\
0~&0~&k_x^2+k_y^2+k_z^2\\
\end{array}\right),
\ee
where 
\be
\mathcal{S}=\left(\begin{array}{ccc}
\frac{k_x}{k_z}~&\frac{k_y}{k_z}~&1\\
-\frac{k_z}{k_x}~&0~&1\\
-\frac{k_y}{k_x}~&1~&0\\
\end{array}\right).
\ee
Introducing the new variables, 
\be
\mathcal{S}\mathcal{Y}\equiv\left(\begin{array}{ccc}
a_x\\
a_y\\
a_z\\
\end{array}\right), 
\ee
and linearly transforming Eq. (\ref{eomAi1}),
\be
\mathcal{S}\left(\ddot{\mathcal{Y}}+c_s^2\mathcal{G}\mathcal{Y}+c_s^2\mathcal{F}\dot{\mathcal{Y}}+c_s^4\mathcal{M}\mathcal{Y}\right)=0,
\ee
and noting that $k_z\dot{a}_x=k_iE_i=\mathcal{O}(h_{ij})$, then up to first order in the gravitational wave amplitude we have the following two coupled differential equations which decouple from the third variable,

\begin{align}\label{eomy}
	 y'' + c_s^2 \tilde{\mathcal{F}}    y' + c_s^2\tilde{k}^2 y +c_s^4  \tilde{\mathcal{M}}  y=0,
\end{align}
where
$y=(a_y,a_z)^T$,  a prime denotes the derivative with respect to $\tau\equiv \frac{\omega t}{2}$, $\tilde{k}^2\equiv 4k^2/\omega^2$,  and 
\begin{align}
	 \tilde{\mathcal{F}} =&
	 h_0\sin 2\tau \begin{pmatrix}
		\frac{\omega^2(k_x+k_y)}{4k_x k^2}  & -\frac{\omega(k_x^2 - k_x k_y +k_z^2)}{2 k_x k^2} \\
		\frac{\omega(-k_x^2 + 2 k_x k_y + k_y^2) }{2 k_x k^2} & \; \frac{-4 k_x k_\epsilon^2- \omega^2 (k_x + k_y)}{4 k_x k^2}
	\end{pmatrix}, \nonumber\\
 \tilde{\mathcal{M}}=&
	h_0 \cos2\tau\begin{pmatrix}
		\frac{2k_\epsilon^2}{\omega^2}+1 + \frac{k_y}{k_x} &\frac{-2k_x+2k_y}{\omega}- \frac{\omega}{2 k_x} \\
		    \frac{2k_y^2}{\omega k_x}-\frac{2 k_x - 4 k_y}{\omega} & \frac{-2\epsilon_{ij}k_ik_j }{\omega^2}-1- \frac{ k_y}{k_x}
	\end{pmatrix},
\end{align}
Note that $a_y\propto F_{xz}$ and $a_z\propto F_{xy}$ are proportional to the gauge field strength and thus gauge invariant.

The equation of motion (\ref{eomy}) has the form of a Mathieu type matrix equation with a friction term. The friction term can be removed via a field rescaling in a similar way to how the Hubble friction term in a scalar field equation can be removed by rescaling the field. The solution of the rescaled variable will then display exponential growth with a Floquet exponent $\mu_k$ in narrow resonance bands of $k$. In terms of the original variables, the exponential growth is modulated by the rescaling function. As shown explicitly in \cite{Craig2} in the case of inflationary reheating, the exponential growth of the solutions trivially extends from the scalar case to the matrix case. 

\begin{figure}[ht]
\subfigure{\includegraphics[width=1.68in]{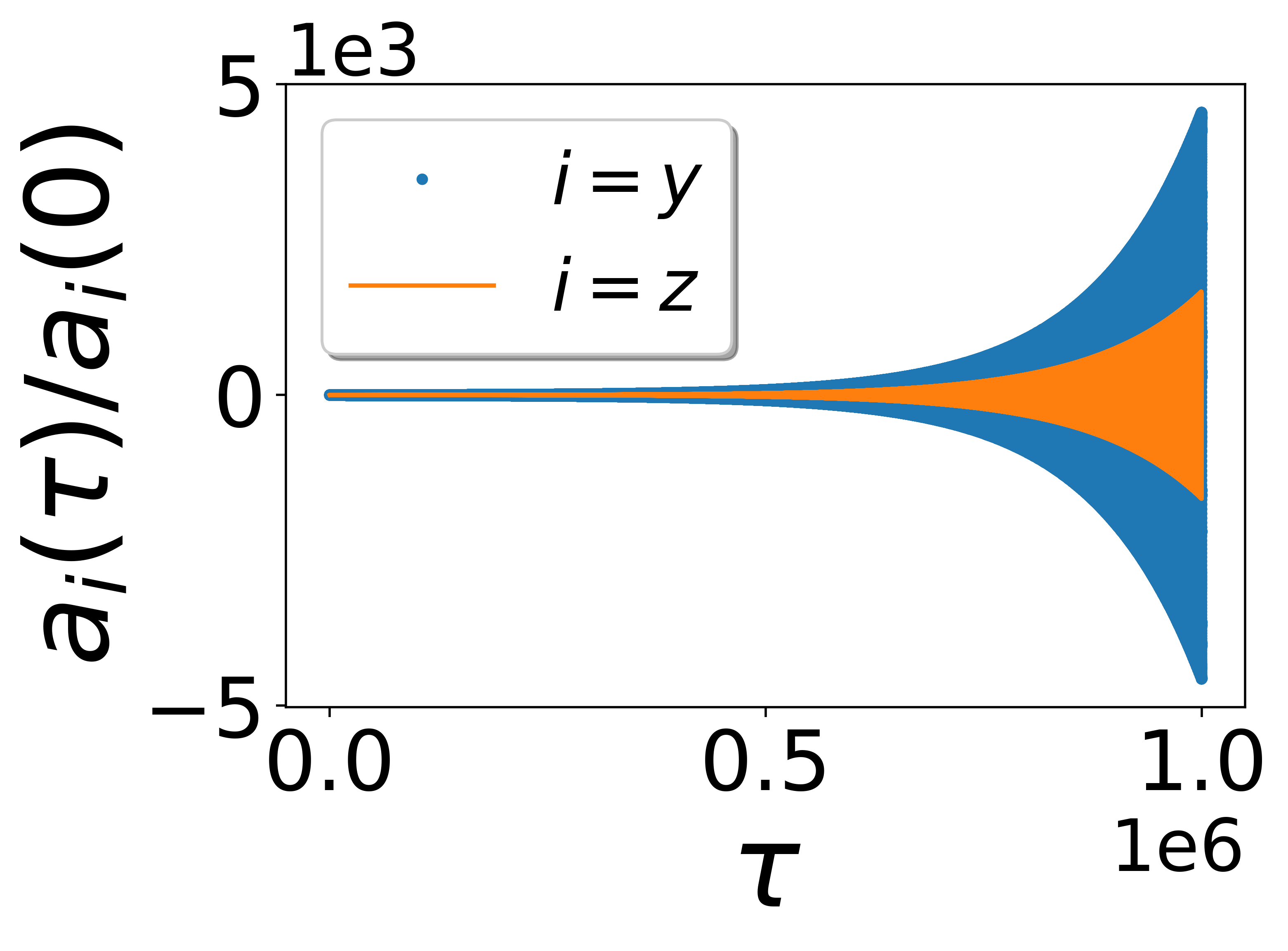}}
\subfigure{\includegraphics[width=1.68in]{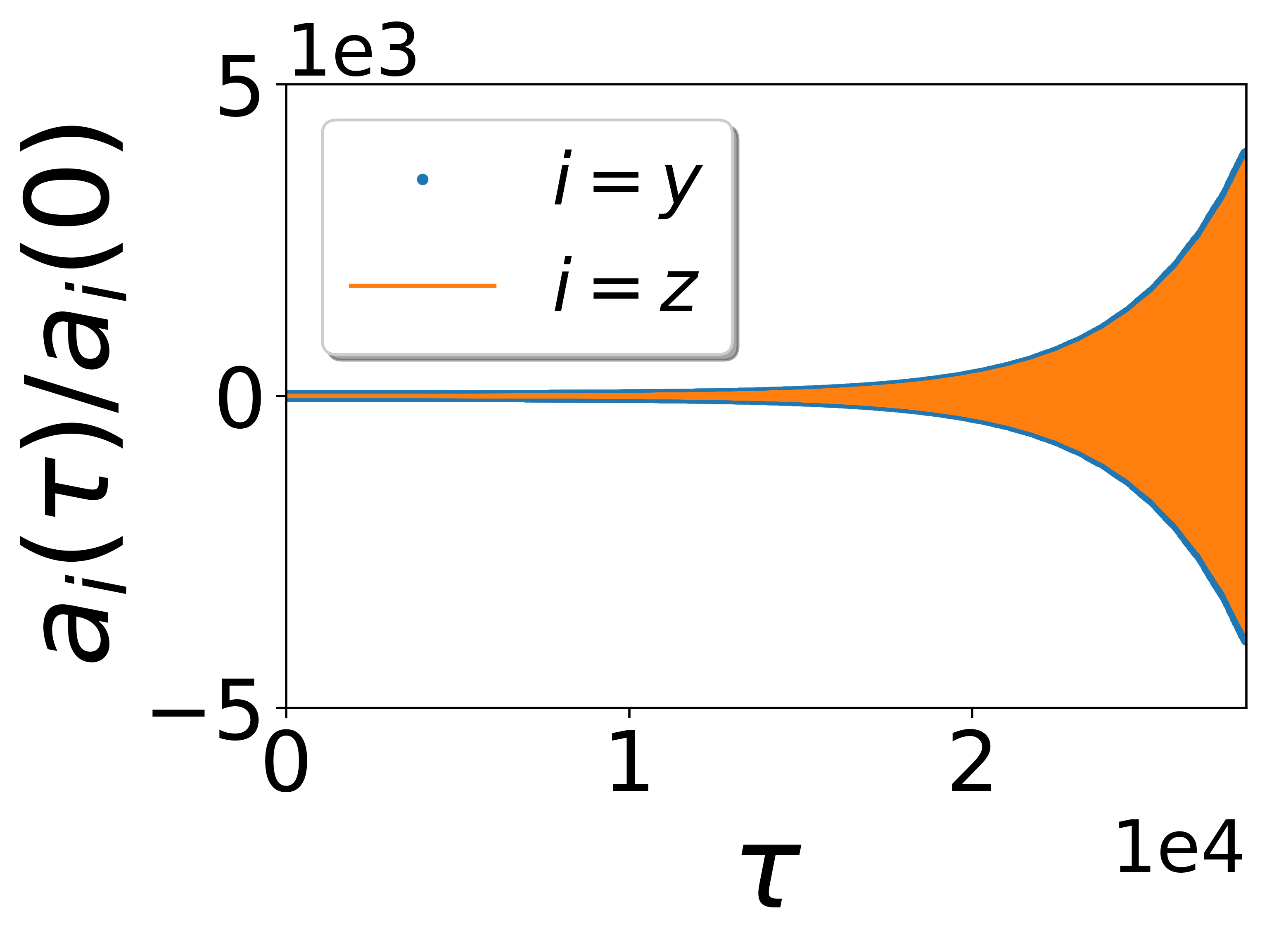}}
\caption{The left figure: the exponential instability of $a_y$ and $a_z$ in the vacuum where $c_s=1$. The resonance occurs at the 2nd band where $A=4$. We set $\tilde{k}_x^2=\tilde{k}_y^2=3/2$ in the numerical plots, and we have adopted a unrealistically large value for  $h_0=0.01$ to reduce the CPU computing time. The initial condition is set to $a_y(0)=a_y'(0)=a_z(0)=a_z'(0)=1$. The right figure: the exponential instability of $a_y$ and $a_z$, where speed of light in the water $c_s=1/1.333$,  $\tilde{k}_x^2=\tilde{k}_y^2=0.388$, and thus we have $A\simeq1$ in the Mathieu equation. We have adopted an unrealistically large value for  $h_0=0.01$ to reduce the CPU computing time. The initial condition is set to $a_y(0)=a_z(0)=1$ and $a_y'(0)=a_z'(0)=0$.}
\label{water}
\end{figure}

We have numerically solved the equation (\ref{eomy}), and the solutions for $a_y$ and $a_z$ are shown in the Fig. \ref{water}, in the first case for propagation in the vacuum ($c_s=1$) and in the second case for propagation in a medium (the value $c_s=1/1.333$ for water was chosen). For $c_s=1$ the resonance occurs only in the second band, while for $c_s=1/1.333$ we have resonance in the first  band. The growth rate in the case of first band resonance is much larger and it takes a much shorter time for the instability to develop.



As is apparent by comparing the two figures, for $c_s^2 < 1$ (the value for water was chosen), the amplication is much stronger (the Floquet exponent is much larger). 
The time scale in the second figure is two orders of magnitude smaller than in the first, and the amplitude at the end of the evolution period is of the same order. The horizontal axis in the graphs is the re-scaled dimensionless time $\tau=\omega t/2$.

Our analysis in this section is based on an un-polarized gravitational standing wave in the flat space-time. The main conclusion also applies to the traveling waves which are of more relevance in various astrophysical phenomena. Namely a traveling gravitational wave can trigger parametric resonance in the photon sector, at the first band in a medium where the refractive index of light differs from unity. In this sense our mechanism is somewhat similar to Cherenkov radiation. However, an essential difference is that the resonant decay rate in our case is proportional to the amount of photon produced at earlier times: namely, it grows exponentially. Moreover, the exponential instability only occurs in a very narrow band in the Mathieu equation, while Cherenkov radiation occurs within a wide frequency range allowed by energy momentum conservation \footnote{Note that the inverse process,
    namely the production of gravitational waves via parametric
    resonance from an oscillating scalar field, does not occur in
    a Minkowski background in a vacuum since the scalar field only enters the
    source term in the gravitational wave equation and not in the
    mass term (see e.e. Eq. 44 of \cite{Zhou:2020kkf}). However, in an
    expanding background, there is the possibility of parametric
    resonance of gravitational waves if the oscillating scalar
    fields lead to small amplitude periodic fluctuations of the
    Hubble expansion rate $H(t)$ superimposed on the regular
    decrease of $H$ (see e.g. \cite{Mesbah}).}. 

In a followup paper \cite{us} the analyses of the various solutions are further developed.\\

\section{ Estimate of the Damping Rate~~~}
In this section we will estimate the decay rate of a wave packet of gravitational waves peaked at frequency $\omega$ with a frequency spread of $\Delta \omega \sim \omega$ due to excitation of electromagnetic fluctuations in a medium with effective speed of light $c_s$. The energy density in gravitational waves is
\be \label{gwenergy}
\rho_{GW} \, \sim \, G^{-1} \omega^2 h_0^2 \, .
\ee
In the semiclassical approximation, we consider vector fields $A_{i}$ initially in their vacuum state, i.e. with an initial amplitude $A_k(t_i) \sim k^{-1/2}$. In this case, the energy density in the produced gauge fields is
\be \label{photonenergy}
\rho_{A} \, \sim \, \Delta \omega \int_{{\cal{P}}} d^2k k^{-1} k^2 e^{2 \mu_k \tau} \, ,
\ee
where the integral runs over the two-dimensional phase space ${\cal{P}}$ of $(k_x, k_y)$ modes which undergo resonance.

For each specific plane wave of frequency $\omega$, resonance occurs for a fixed value of $k_z$, namely $k_z = \omega / 2$, and for a band of $(k_x, k_y)$ with width $r_{max}^2-r_{min}^2\simeq \frac{q\omega^2}{4 c_s^2}$ and radius $r$ determined by
\be \label{res2}
r^2 \, = \, k_x^2 + k_y^2 \, \simeq \frac{1 - c_s^2}{4c_s^2} \omega^2 \, .
\ee

These two equations determine the range of values of $(k_x, k_y)$ for which $A_k = 1$ modulo $q$. Thus, in (\ref{photonenergy}) we make the approximations of replacing the modulus $k$ by $\omega / 2$ and taking $\mu_k$ to be independent of $k$. Inserting the eq. (\ref{res2}), $\delta \omega \sim \omega$ and the value $q \sim c_s^2 (1-c_s^2) h_0$ we obtain
\begin{equation}
\rho_A \sim \omega^4 c_s^4 (1-c_s^2) h_0 e^{2 \mu \tau}.
\end{equation}
The decay rate of the gravitational wave amplitude $h_0$ can then be determined by equating the energy gain in $\rho_A$ with the energy loss in $\rho_{GW}$. Neglecting the time dependence of $h_0$ in $\rho_A$ (the time dependence is dominated by the Floquet term and including the time dependence of $h_0$ would yield only a higher order correction) yields
\be
\log (h_0 )^\prime \sim - G \omega^2 c_s^6 (1-c_s^2)^2 e^{2 \mu \tau}.
\ee
Thus, we see that the decay rate of $h_0$ on the gravitational wave oscillation time scale is suppressed by  $G\omega^2 $ and also by the factor $(1 - c_s^2)^2$.\\



\section{ Conclusions and Discussion~~~} 
We have shown that gravitational waves can be damped by exciting a parametric resonance instability of the electromagnetic gauge field. In vacuum, the resonance is very weak since the resonant modes lie in the second resonance band. In a medium in which electromagnetic waves travel with a speed smaller than $1$, on the other hand, the resonance is in the first band and hence stronger. We have estimated the decay rate which a wavepacket of gravitational waves undergoes.


The analysis is based on a single gravitational wave with fixed frequency. The extension to several gravitational wave modes is straightforward. As to be expected from the general theory of Floquet instability and also studied explicitly for inflationary reheating in \cite{Craig2}, the instability remains, and the Floquet exponent for a fixed value of $k_z = \omega / 2$ is boosted if gravitational waves of different frequencies are added. This will also be discussed in \cite{us}.

The conversion of gravitational waves into plasma waves has been studied in the literature focusing on linear resonant conversion \cite{Chen:1994ch} or the non-linear interaction of two plasma and one gravitational wave \cite{Brodin:1998xj,PhysRevE.62.8493,Forsberg:2006fu,Brodin:2009yy}, in the presence of strong background magnetic fields. Our analysis fits nicely into this area providing a new conversion process with the same order of magnitude for the growth parameter as for the three wave interaction \cite{PhysRevE.62.8493}, without requiring a strong background magnetic field to exist, provided that the plasma mass is sufficiently small compared to the frequency of the gravitational wave,
$ m_{{\rm{plasma}}}^2 \, < \, (1 - c_s^2) \omega^2/4 $.

Our result is a first step in the direction of investigating possible implications of gravitational wave conversion via parametric resonance in cosmology and astrophysics. The biggest challenge in finding straightforward applications is to achieve the necessary conditions that lead to a non-negligible conversion rate: namely a refractive index sufficiently larger than $1$ in a context where 
there is enough time for the instability to develop. In a black hole binary, for instance, the orbital decay is faster than the time required for a non-negligible conversion, while the refractive index in the accretion disk is generally not large enough. In the early universe, during radiation domination, the refractive index is indeed significant, and a field redefined in order to incorporate the background expansion satisfies a Mathieu equation with $q\sim h_0/\omega^2$. It would be interesting to carefully investigate the possibility of suppression of B-modes in the Cosmic Microwave Background if the instability is well developed until matter-radiation equality. Another potential application consists in a novel type of gravitational wave detector, in which gravitational waves turn into possibly detectable electromagnetic waves whose amplitude grows as $\exp\left(\epsilon h_0\omega t\right)$ due to the exponential instability induced by parametric resonance, where $h_0$ and $\omega$ are the amplitude and frequency of gravitational waves respectively, and $\epsilon$ is an order one constant depending on the relation between the momentum of electromagnetic waves and gravitational waves. Given the amplitude of gravitational waves, the electromagnetic signals grow faster for high frequency gravitational waves. It remains a challenge to detect the high frequency gravitational waves with natural origin \cite{Vagnozzi:2022qmc}. Nevertheless, there are already some ideas about lab generation of high frequency gravitational waves \cite{ADL2023, Gorelik, Pustovoit}.  Finally, we shall mention that the methodology developed in this work can be applied to investigate the inverse process, namely the amplification of gravitational waves due to parametric resonance. We leave the above-mentioned possibilities for future work.

\noindent 

\section*{Declaration of competing interest}
The authors declare that they have no known competing financial interests or personal relationships that could have appeared to influence the work reported in this paper. 
\section*{Data availability}
No data was used for the research described in the article.

\section*{Acknowledgement}

We would like to thank Yifu Cai, Bryce Cyr, Charles Dalang, Misao Sasaki, Yi Wang, Sunny Vagnozzi and Lidiia Zadorozhna for useful discussions. C.L. and P.C.M.D. are supported by the grant No. UMO-2018/30/Q/ST9/00795 from the National Science Centre, Poland. A.G. receives support from the grant No. UMO-2021/40/C/ST9/00015 from the National Science Centre, Poland.
 The research at McGill is supported in
part by funds from NSERC and from the Canada Research Chair
program. RB is grateful for hospitality of the Institute for
Theoretical Physics and the Institute for Particle Physics and
Astrophysics of the ETH Zurich.



\begin{thebibliography}{99}

\bibitem{Landau} L. Landau and E. M. Lifshitz, {\it Mechanics} (3rd Edition)(Elsevier, Oxford, 1976)

\bibitem{Arnold} V. Arnold, {\it Mathematical Methods of Classical Mechanics} (Springer, Berlin, 1978).

\bibitem{Floquet}
N. W. McLachlan, {\it Theory and Applications of Mathieu
Functions} (Oxford Univ. Press, Clarendon, 1947).

\bibitem{TB}
J.~H.~Traschen and R.~H.~Brandenberger,
``Particle Production During Out-of-equilibrium Phase Transitions,''
Phys. Rev. D \textbf{42}, 2491-2504 (1990)
doi:10.1103/PhysRevD.42.2491

\bibitem{DK}
A.~D.~Dolgov and D.~P.~Kirilova,
``ON PARTICLE CREATION BY A TIME DEPENDENT SCALAR FIELD,''
Sov. J. Nucl. Phys. \textbf{51}, 172-177 (1990)
JINR-E2-89-321.

\bibitem{RHrev}
R.~Allahverdi, R.~Brandenberger, F.~Y.~Cyr-Racine and A.~Mazumdar,
``Reheating in Inflationary Cosmology: Theory and Applications,''
Ann. Rev. Nucl. Part. Sci. \textbf{60}, 27-51 (2010)
doi:10.1146/annurev.nucl.012809.104511
[arXiv:1001.2600 [hep-th]].

\bibitem{Karouby} 
M.~A.~Amin, M.~P.~Hertzberg, D.~I.~Kaiser and J.~Karouby,
``Nonperturbative Dynamics Of Reheating After Inflation: A Review,''
Int. J. Mod. Phys. D \textbf{24}, 1530003 (2014)
doi:10.1142/S0218271815300037
[arXiv:1410.3808 [hep-ph]].

\bibitem{KLS1}
L.~Kofman, A.~D.~Linde and A.~A.~Starobinsky,
``Reheating after inflation,''
Phys. Rev. Lett. \textbf{73}, 3195-3198 (1994)
doi:10.1103/PhysRevLett.73.3195
[arXiv:hep-th/9405187 [hep-th]].

\bibitem{STB}
Y.~Shtanov, J.~H.~Traschen and R.~H.~Brandenberger,
``Universe reheating after inflation,''
Phys. Rev. D \textbf{51}, 5438-5455 (1995)
doi:10.1103/PhysRevD.51.5438
[arXiv:hep-ph/9407247 [hep-ph]].

\bibitem{KLS2}
L.~Kofman, A.~D.~Linde and A.~A.~Starobinsky,
``Towards the theory of reheating after inflation,''
Phys. Rev. D \textbf{56}, 3258-3295 (1997)
doi:10.1103/PhysRevD.56.3258
[arXiv:hep-ph/9704452 [hep-ph]].

\bibitem{Felder}
P.~B.~Greene and L.~Kofman,
``Preheating of fermions,''
Phys. Lett. B \textbf{448}, 6-12 (1999)
doi:10.1016/S0370-2693(99)00020-9
[arXiv:hep-ph/9807339 [hep-ph]].


\bibitem{Cai:2018tuh}
Y.~F.~Cai, X.~Tong, D.~G.~Wang and S.~F.~Yan,
Phys. Rev. Lett. \textbf{121}, no.8, 081306 (2018)
[arXiv:1805.03639 [astro-ph.CO]].

\bibitem{Cai:2020ovp}
Y.~F.~Cai, C.~Lin, B.~Wang and S.~F.~Yan,
Phys. Rev. Lett. \textbf{126} (2021) no.7, 071303
doi:10.1103/PhysRevLett.126.071303
[arXiv:2009.09833 [gr-qc]].



\bibitem{Craig2}
V.~Zanchin, A.~Maia, Jr., W.~Craig and R.~H.~Brandenberger,
``Reheating in the presence of inhomogeneous noise,''
Phys. Rev. D \textbf{60}, 023505 (1999)
doi:10.1103/PhysRevD.60.023505
[arXiv:hep-ph/9901207 [hep-ph]].

\bibitem{Furstenberg}
R. Carmona and J. Lacroix, {\it Spectral theory of random Schroedinger operators} 
(Birkhaeuser, Boston. 1990). See p 198 for the proof of existence of the Floquet exponents and p 200 for the proof of the Furstenberg theorem.

\bibitem{Craig1}
V.~Zanchin, A.~Maia, Jr., W.~Craig and R.~H.~Brandenberger,
``Reheating in the presence of noise,''
Phys. Rev. D \textbf{57}, 4651-4662 (1998)
doi:10.1103/PhysRevD.57.4651
[arXiv:hep-ph/9709273 [hep-ph]].

\bibitem{Craig3}
R.~Brandenberger and W.~Craig,
``Towards a New Proof of Anderson Localization,''
Eur. Phys. J. C \textbf{72}, 1881 (2012)
doi:10.1140/epjc/s10052-012-1881-9
[arXiv:0805.4217 [hep-th]].

\bibitem{Creminelli:2019nok}
P.~Creminelli, G.~Tambalo, F.~Vernizzi and V.~Yingcharoenrat,
JCAP \textbf{10} (2019), 072
doi:10.1088/1475-7516/2019/10/072
[arXiv:1906.07015 [gr-qc]].


\bibitem{us}
R. Brandenberger, P. C. M. Delgado, A. Ganz, and C. Lin, in preparation.

\bibitem{Chen:1994ch}
P.~Chen,
Phys. Rev. Lett. \textbf{74} (1995), 634-637
[erratum: Phys. Rev. Lett. \textbf{74} (1995), 3091]
doi:10.1103/PhysRevLett.74.634

\bibitem{Brodin:1998xj}
G.~Brodin and M.~Marklund,
Phys. Rev. Lett. \textbf{82} (1999), 3012-3015
doi:10.1103/PhysRevLett.82.3012
[arXiv:astro-ph/9810128 [astro-ph]].
\bibitem{Forsberg:2006fu}
M.~Forsberg, G.~Brodin, M.~Marklund, P.~K.~Shukla and J.~Moortgat,
Phys. Rev. D \textbf{74} (2006), 064014
doi:10.1103/PhysRevD.74.064014
[arXiv:gr-qc/0606072 [gr-qc]].
\bibitem{Brodin:2009yy}
G.~Brodin, M.~Forsberg, M.~Marklund and D.~Eriksson,
J. Plasma Phys. \textbf{76} (2010), 345
doi:10.1017/S0022377809990535
[arXiv:0911.2190 [gr-qc]].

\bibitem{PhysRevE.62.8493}
 M.~Servin, G.~Brodin, M.~Bradley and M.~Marklund,
  Phys. Rev. E. \textbf{62} (2000), doi:10.1103/PhysRevE.62.8493
 [arXiv::physics/9910029v2].
\bibitem{Zhou:2020kkf}
Z.~Zhou, J.~Jiang, Y.~F.~Cai, M.~Sasaki and S.~Pi,
Phys. Rev. D \textbf{102} (2020) no.10, 103527
doi:10.1103/PhysRevD.102.103527
[arXiv:2010.03537 [astro-ph.CO]].
 \bibitem{Mesbah} M. Alsarraj and R. Brandenberger, arXiv:2103.07684.

\bibitem{Vagnozzi:2022qmc}
S.~Vagnozzi and A.~Loeb,
Astrophys. J. Lett. \textbf{939} (2022) no.2, L22
doi:10.3847/2041-8213/ac9b0e
[arXiv:2208.14088 [astro-ph.CO]].

\bibitem{ADL2023}
S. Akama, P. C. M. Delgado, C. Lin, in preparation.

\bibitem{Gorelik}
V. S. Gorelik et al, J. Phys.: Conf. Ser. 1051 012001 (2018)
doi:10.1088/1742-6596/1051/1/012001.

\bibitem{Pustovoit}
V. I. Pustovoit et al, J. Phys.: Conf. Ser. 1348 012008 (2019)
doi:10.1088/1742-6596/1348/1/012008.


\end{thebibliography}
\end{document}